\documentclass[10pt,twocolumn,letterpaper]{article}

\usepackage{cvpr}
\usepackage{times}
\usepackage{epsfig}
\usepackage{graphicx}
\usepackage{amsmath}
\usepackage{amssymb}
\usepackage{mathrsfs}
\usepackage{slashbox}
\usepackage{dsfont}
\usepackage{booktabs}
\usepackage{subcaption}
\usepackage{caption}
\usepackage{dblfloatfix}
\usepackage{mathrsfs}
\usepackage{xcolor}
\usepackage{colortbl}
\usepackage{multirow}
\usepackage{paralist}
\usepackage{marvosym}
\usepackage[export]{adjustbox}
\newcommand{\ch}{\checkmark}
\definecolor{sh_gray}{rgb}{0.84,0.84,0.84}
\definecolor{sh_gray2}{rgb}{1,0.89,0.75}
\definecolor{color3}{rgb}{0.95,0.95,0.95}
\definecolor{color4}{rgb}{0.96,0.96,0.86}
\definecolor{color5}{rgb}{0.90,0.90,0.90}

\newlength{\Oldarrayrulewidth}

\usepackage[pagebackref=true,breaklinks=true,letterpaper=true,colorlinks,linkcolor=blue,citecolor=blue,urlcolor=blue,bookmarks=false]{hyperref}

\cvprfinalcopy 

\usepackage{cuted}
\usepackage{capt-of}

\title{\vspace{-0.7em}CycleISP: Real Image Restoration via Improved Data Synthesis}

\begin{document}


\author{Syed Waqas Zamir$^{1}$ \quad Aditya Arora$^{1}$ \quad Salman Khan$^{1}$ \quad Munawar Hayat$^{1}$ \\ 
Fahad Shahbaz Khan$^{1}$  \quad Ming-Hsuan Yang$^{2,3}$ \quad Ling Shao$^{1}$ \\
$^1$Inception Institute of Artificial Intelligence,  UAE\\
$^2$University of California, Merced \quad $^3$Google Research 
}

\maketitle

\begin{abstract}\vspace{-1.0em}
The availability of large-scale datasets has helped unleash the true potential of deep convolutional neural networks (CNNs).
However, for the single-image denoising problem, capturing a real dataset is an unacceptably expensive and cumbersome procedure. 
Consequently, image denoising algorithms are mostly developed and evaluated on synthetic data that is usually generated with a widespread assumption of additive white Gaussian noise (AWGN). 
While the CNNs achieve impressive results on these synthetic datasets, they do not perform well when applied on real camera images, as reported in recent benchmark datasets. 
This is mainly because the AWGN is not adequate for modeling the real camera noise which is signal-dependent and heavily transformed by the camera imaging pipeline. 
In this paper, we present a framework that models camera imaging pipeline in forward and reverse directions. 
It allows us to produce any number of realistic image pairs for denoising both in RAW and sRGB spaces. 
By training a new image denoising network on realistic synthetic data, we achieve the state-of-the-art performance on real camera benchmark datasets. The parameters in our model are $\sim$5 times lesser than the previous best method for RAW denoising.  
Furthermore, we demonstrate that the proposed framework generalizes beyond image denoising problem e.g., for color matching in stereoscopic cinema.
The source code and pre-trained models are available at \url{https://github.com/swz30/CycleISP}.
\end{abstract}
\vspace{-0.3cm}

\begin{figure}[t]
  \begin{center}
    \begin{tabular}{cc}\hspace{-4mm}
      \includegraphics[width=0.224\textwidth]{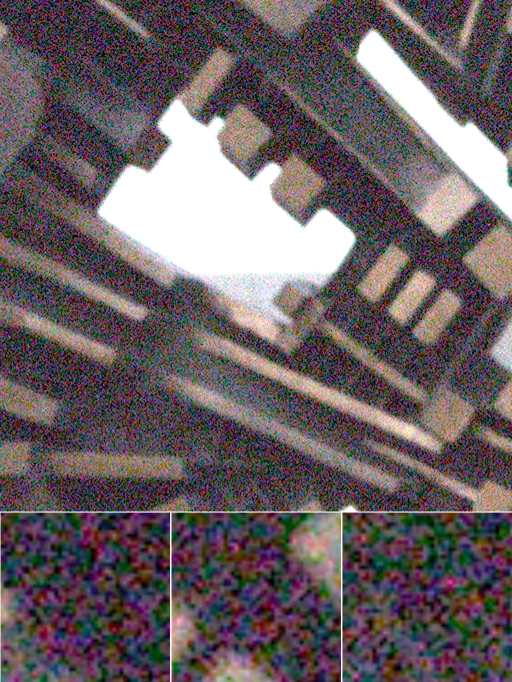}&\hspace{-7mm}
      \includegraphics[width=0.224\textwidth]{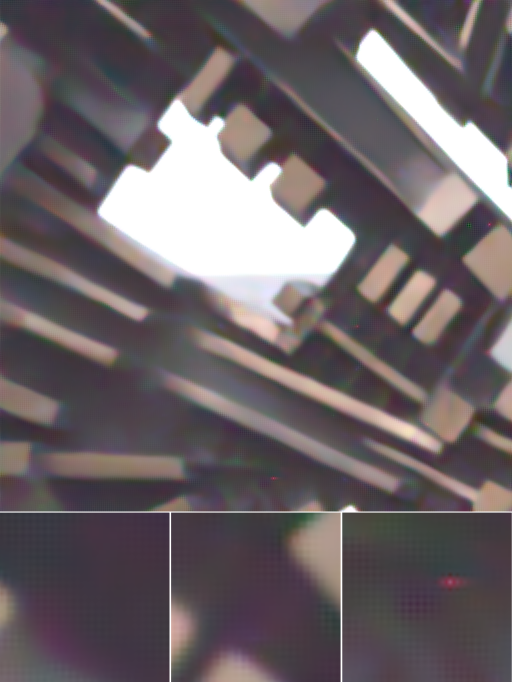}\hspace{-6mm}\\
      (a) Noisy Input & \hspace{-5mm} (b) N3NET \cite{N3Net}  \\
      PSNR(RAW) / PSNR(sRGB) & \hspace{-5mm} 38.24 dB / 32.42 dB \\
      \hspace{-4mm}
      \includegraphics[width=0.224\textwidth]{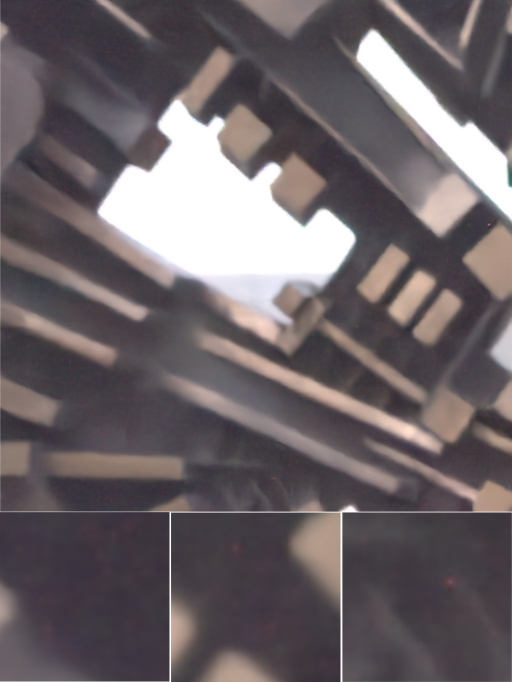}&\hspace{-7mm}
      \includegraphics[width=0.224\textwidth]{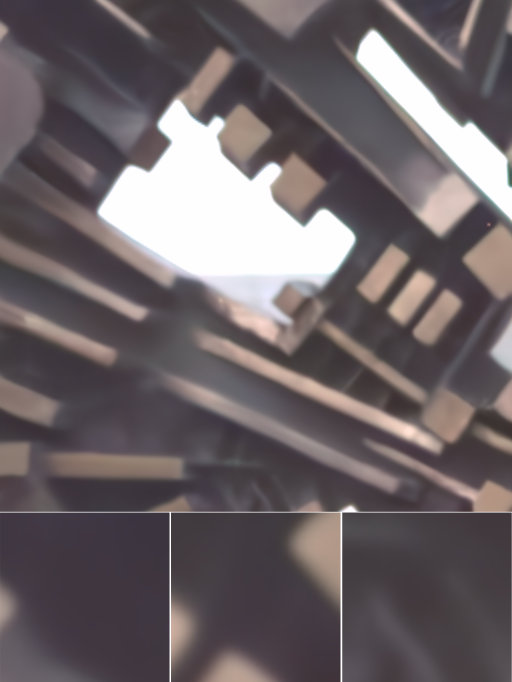}\hspace{-6mm}\\
      (c) UPI \cite{Brooks2019}  & \hspace{-5mm} (d) Ours \\
      37.37 dB / 35.49 dB & \hspace{-5mm}  \textbf{40.44 dB / 36.16 dB}\\
    \end{tabular}
  \end{center}\vspace{-1.8em}
 \caption{Denoising a real camera image from DND dataset  \cite{dnd}.
 Our model is effective in removing real noise, especially the low-frequency chroma and defective pixel noise.} \vspace{-0.8em} 
 \vspace{-2mm}
\label{Fig:teaser}
\end{figure}

\vspace{-0.5em}
\section{Introduction}

High-level computer vision tasks, such as image classification, object detection and segmentation have witnessed significant progress due to deep CNNs~\cite{khan2018guide}. 
The major driving force behind the success of CNNs is the availability of large-scale datasets \cite{imagenet,mscoco}, containing hundreds of thousands of annotated images. 
However, for \emph{low-level} vision problems (image denoising, super-resolution, 
deblurring, etc.), collecting even small datasets is extremely challenging and non-trivial. 
For instance, the typical procedure to acquire noisy paired data is to take multiple noisy images of the same scene and generate clean ground-truth image by pixel-wise averaging.
In practice, spatial pixels misalignment, color and brightness mismatch is inevitable due to changes in lighting conditions and camera/object motion. 
Moreover, this expensive and cumbersome exercise of acquiring image pairs needs to be repeated with different camera sensors, as they exhibit different noise characteristics.   

Consequently, single image denoising is mostly performed in synthetic settings: take a large set of clean sRGB images and add synthetic noise to generate their noisy versions. 
On synthetic datasets, existing deep learning based denoising models yield impressive results, but they exhibit poor generalization to real camera data as compared to conventional methods \cite{NLM,BM3D}. 
This trend is also demonstrated in recent benchmarks \cite{sidd,dnd}. %
Such behavior stems from the fact that deep CNNs are trained on synthetic data that is usually generated with the Additive White Gaussian Noise (AWGN) assumption. 
Real camera noise is fundamentally different from AWGN, thereby causing a major challenge for deep CNNs \cite{bertalmio2018denoising,foi2009noisemodel,foi2008noisemodel}. 

In this paper, we propose a synthetic data generation approach that can produce realistic noisy images both in RAW and sRGB spaces. 
The main idea is to inject noise in the RAW images obtained with our learned device-agnostic transformation rather than in the sRGB images directly. 
The key insight behind our framework is that the real noise present in sRGB images is convoluted by the series of steps performed in a regular image signal processing (ISP) 
pipeline \cite{bertalmio2018denoising, Ramanath2005}.
Therefore, modeling real camera noise in sRGB is an inherently difficult task as compared to RAW sensor data \cite{lebrun2012secrets}. As an example, noise at the RAW sensor space is signal-dependent; after demosaicking, it becomes spatio-chromatically correlated; and after passing through the rest of the pipeline, its probability distribution not necessarily remains Gaussian \cite{seybold2014noise}. 
This implies that the camera ISP heavily transforms the sensor noise, and therefore more sophisticated models that take into account the influence of imaging pipeline are needed to synthesize realistic noise than uniform AWGN model \cite{sidd,ghimpecteanu2016local,dnd}.

In order to exploit the abundance and diversity of sRGB photos available on the Internet, the main challenge with the proposed synthesis approach is how to transform them back to RAW measurements.
Brooks \etal \cite{Brooks2019} present a technique that inverts the camera ISP, step-by-step, and thereby allows conversion from sRGB to RAW data.
However, this approach 
requires prior information about the target camera device (e.g., color correction matrices and white balance gains), which makes it specific to a given device and therefore lacks in generalizability. 
Furthermore, several operations in a camera pipeline are proprietary and such black boxes are very difficult to reverse engineer. 
To address these challenges, in this paper we propose a CycleISP framework that converts sRGB images to RAW data, and then back to sRGB images, without requiring any knowledge of camera parameters. 
This property allows us to synthesize any number of clean and realistic noisy image pairs in both RAW and sRGB spaces. Our main contributions are:

\begin{compactitem}
    \item Learning a device-agnostic transformation, called CycleISP, that allows us to move back and forth between sRGB and RAW image spaces.
    \item Real image noise synthesizer for generating clean/noisy paired data in RAW and sRGB spaces. 
    \item A deep CNN with dual attention mechanism that is effective in a variety of tasks: learning CycleISP, synthesizing realistic noise, and image denoising.
    \item Algorithms to remove noise from RAW and sRGB images, setting new state-of-the-art on real noise benchmarks of DND \cite{dnd} and SIDD \cite{sidd} (see Fig.~\ref{Fig:teaser}). Moreover, our denoising network  has much fewer parameters (2.6M) than the previous best model (11.8M) \cite{Brooks2019}.
    \item CycleISP framework generalizes beyond denoising, we demonstrate this via an additional application \ie, color matching in stereoscopic cinema.
\end{compactitem}

\section{Related Work}
The presence of noise in images is inevitable, irrespective of the acquisition method; now more than ever, when majority of images come from smartphone cameras having small sensor size but large resolution. 
Single-image denoising is a vastly researched problem in the computer vision and image processing community, with early works dating back to 1960's \cite{bertalmio2018denoising}.
Classic methods on denoising are mainly based on the following two principles. (1) Modifying transform coefficients using the DCT \cite{yaroslavsky1996local}, wavelets \cite{donoho1995noising,simoncelli1996noise}, etc. (2) Averaging neighborhood values: in all directions using Gaussian kernel, in all directions only if pixels have similar values \cite{smith1997susan,tomasi1998bilateral} and along contours \cite{perona1990scale,rudin1992nonlinear}.

While these aforementioned methods provide satisfactory results 
in terms of image fidelity metrics and visual quality, the Non-local Means (NLM) algorithm of Buades \etal \cite{NLM} makes significant advances in denoising.
The NLM method exploits the redundancy, or self-similarity \cite{efros1999texture} present in natural images.
For many years the patch-based methods yielded comparable results, thus prompting studies \cite{chatterjee2009denoising,chatterjee2010fundamental,levin2011natural} to investigate whether we reached the theoretical limits of denoising performance. 
Subsequently, Burger \etal \cite{MLP} train a  simple Multi-Layer Perceptron (MLP) on a large synthetic noise dataset.
This method performs well against previous sophisticated algorithms. 
Several recent methods use deep CNNs \cite{RIDNet,Brooks2019,Gharbi2016,CBDNet,N3Net,DnCNN,FFDNetPlus,ntire2019denoising} and demonstrate promising denoising performance. 

\begin{figure*}[t!]
\begin{center}
 \includegraphics[width=0.85\linewidth]{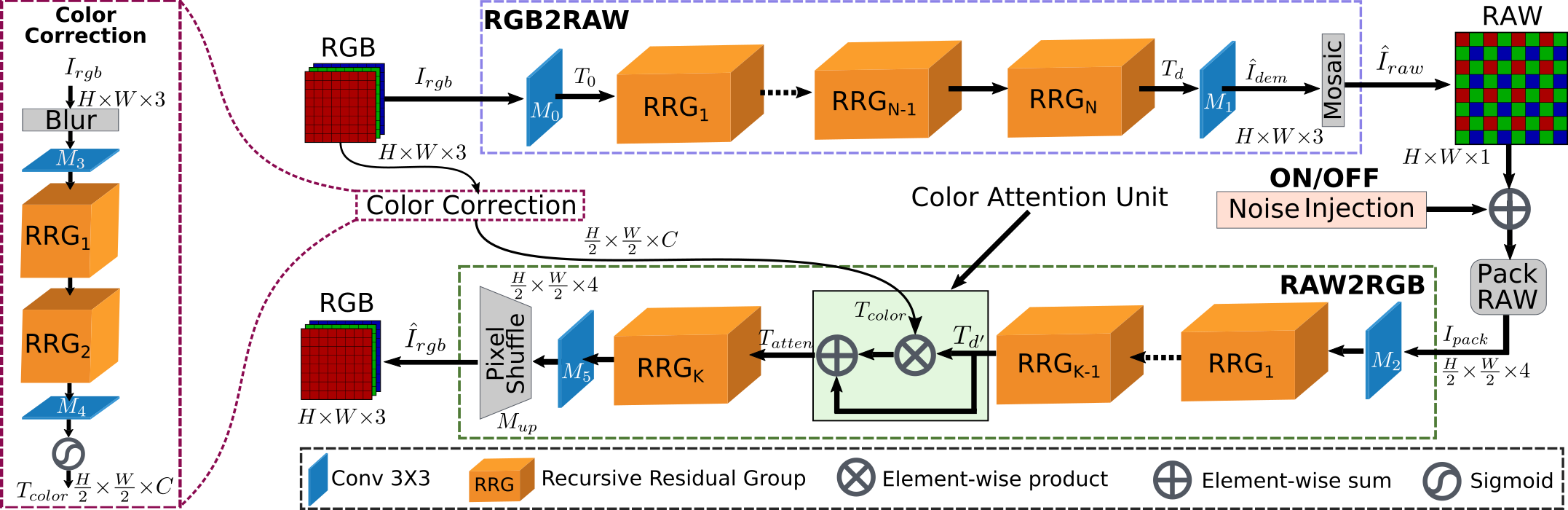}  
\end{center}\vspace{-1.6em}
    \caption{Our CycleISP models the camera imaging pipeline in both directions. It comprises two main branches: RGB2RAW and RAW2RGB. The RGB2RAW branch converts sRGB images to RAW measurements, whereas the RAW2RGB branch transforms RAW data to sRGB images. The auxiliary color correction branch provides explicit color attention to RAW2RGB network. The noise injection module is switched OFF while training the CycleISP (Section~\ref{sec:cycleisp}), and switched ON when synthesizing noise data (Section~\ref{sec:finetune noisy}).}\vspace{-2.5mm}
    \label{Fig:cycleisp}
\end{figure*}

Image denoising can be applied to RAW or sRGB data.
However, capturing diverse large-scale real noise data is a prohibitively expensive and tedious procedure, consequently leaving us to study denoising in synthetic settings. 
The most commonly used noise model for developing and evaluating image denoising is AWGN. 
As such, algorithms that are designed for AWGN cannot effectively remove noise from real images, as reported in recent benchmarks \cite{sidd,dnd}. 
A more accurate model for real RAW sensor noise contains both the signal-dependent noise component (the shot noise), and the  signal-independent additive  Gaussian component (the read noise) \cite{foi2009noisemodel,foi2007noisemodel,foi2008noisemodel}. 
The camera ISP transforms RAW sensor noise into a complicated form (spatio-chromatically correlated and not necessarily Gaussian).
Therefore, estimating a noise model for denoising in sRGB space requires careful consideration of the influence of ISP. 
In this paper, we present a framework that is capable of synthesizing realistic noise data for training CNNs to effectively remove noise from RAW as well as sRGB images.

\section{CycleISP}
\label{sec:cycleisp}
To 
synthesize realistic noise datasets, we use a two-stage scheme in this work.
First, we develop a framework that models the camera ISP both in forward and reverse directions, 
hence the name CycleISP. 
Second, using CycleISP, we synthesize realistic noise datasets for the tasks of RAW denoising and sRGB image denoising.
In this section, we only describe our CycleISP framework that models the camera ISP as a deep CNN system. 
Fig.~\ref{Fig:cycleisp} shows the modules of the CycleISP model: (a) RGB2RAW network branch, and (b) RAW2RGB network branch.
In addition, we introduce an auxiliary color correction network branch that provides explicit color attention to the RAW2RGB network in order to correctly recover the original sRGB image.   

The noise injection module in Fig.~\ref{Fig:cycleisp} is only required when synthesizing noisy data (Section~\ref{sec:finetune noisy}), and thus we keep it in the `OFF' state while learning CycleISP. 
The training process of CycleISP is divided in two steps: the RGB2RAW and RAW2RGB networks are first independently trained, and then joint fine-tuning is performed. 
Next, we present details of different branches of CycleISP. 
Note that we use RGB instead of sRGB to avoid notation clutter.

\subsection{RGB2RAW Network Branch}
Digital cameras apply a series of operations on RAW sensor data in order to generate the monitor-ready sRGB images \cite{Ramanath2005}. 
Our RGB2RAW network branch aims to invert the effect of camera ISP. 
In contrast to the \emph{unprocessing} technique of \cite{Brooks2019}, the RGB2RAW branch does not require any camera parameters.  

Given an input RGB image $\mathbf{I}_{rgb} \in \mathbb{R}^{H\times W \times 3}$, the RGB2RAW network first extracts low-level features $T_0\in \mathbb{R}^{H\times W \times C}$ using a convolutional layer $M_0$ as: $T_0 = M_0(\mathbf{I}_{rgb}).$ 
Next, we pass the low-level feature maps $T_0$ through $N$ recursive residual groups (RRGs) to extract deep features $T_d \in \mathbb{R}^{H\times W \times C}$ as:
\begin{align}
T_d = RRG_{N}\left(...(RRG_1(T_0))\right),
\label{Eq:rrg}
\end{align}
where each RRG contains multiple dual attention blocks, as we shall see in Section~\ref{sec:rrg}. 

We then apply the final convolution operation $M_1$ to the features $T_d$ and obtain the demosaicked image $\hat{I}_{dem} \in \mathbb{R}^{H\times W \times 3}$.
We deliberately set the number of output channels of $M_1$ layer to three rather than one in order to preserve as much structural information of the original image as possible. 
Moreover, we empirically found that it helps the network to learn the mapping from sRGB to RAW faster and more accurately. 
At this point, the network is able to invert the effects of tone mapping, gamma correction, color correction, white balance, and other transformations, and provide us with the image $\hat{I}_{dem}$ whose values are linearly related to the scene radiance. 
Finally, in order to generate the mosaicked RAW output $\hat{\mathbf{I}}_{raw}\in \mathbb{R}^{H\times W \times 1}$, the Bayer sampling function $f_{Bayer}$ is applied to $\hat{I}_{dem}$ that omits two color channels per pixel according to the  Bayer pattern: 
\begin{align}
\hat{\mathbf{I}}_{raw} = f_{bayer}(M_1(T_d)).
\label{Eq:raw}
\end{align}

The RGB2RAW network is optimized using the $L_1$ loss in linear and log domains as:
\begin{equation}
\label{Eq:loss rgb2raw}
\begin{split}
\mathcal{L}&_{{s\rightarrow r}}(\hat{\mathbf{I}}_{raw},\mathbf{I}_{raw})
=
\left \| \hat{\mathbf{I}}_{raw} - \mathbf{I}_{raw}  \right \|_{1} 
\\
&
+ \left \| \log(\text{max}(\hat{\mathbf{I}}_{raw},\epsilon)) - \log(\text{max}(\mathbf{I}_{raw},\epsilon))\right \|_{1} ,
\end{split}
\end{equation}
where $\epsilon$ is a small constant for numerical stability, and $\mathbf{I}_{raw}$ is the ground-truth RAW image. 
Similar to \cite{eilertsen2017hdr}, the log loss term is added to enforce approximately equal treatment for all the image values; otherwise the network dedicates more attention to recovering the highlight regions.

\subsection{RAW2RGB Network Branch}
While the ultimate goal of RAW2RGB network is to  generate synthetic realistic noise data for the sRGB image denoising problem, in this section we first describe how we can map \emph{clean} RAW images to \emph{clean} sRGB images (leaving the noise injection module `OFF' in Fig.~\ref{Fig:cycleisp}).

Let $\mathbf{I}_{raw}$ 
and $\hat{\mathbf{I}}_{rgb}$ be
the input and output of the RAW2RGB network.
First, in order to restore translation invariance and reduce computational cost, we pack the $2{\times}2$ blocks of $\mathbf{I}_{raw}$ into four channels (RGGB) and thus reduce the image resolution by half \cite{Brooks2019,Chen2018,Gharbi2016}. 
Since the input RAW data may come from different cameras having different Bayer patterns, we ensure the channel order of the packed image to be RGGB by applying the Bayer pattern unification technique \cite{liu2019learning}. 
Next, a convolutional layer $M_2$ followed by ${K-1}$ RRG modules encode the packed RAW image $I_{pack}\in \mathbb{R}^{\frac{H}{2}{\times}\frac{W}{2}{\times} 4}$ into a deep feature tensor $T_{d'}\in \mathbb{R}^{\frac{H}{2}{\times}\frac{W}{2}{\times} C}$ as:
\begin{align}
T_{d'} = RRG_{K-1}(...(RRG_1(M_2(\text{Pack}(\mathbf{I}_{raw}))))).
\label{Eq:raw2rgb DF tensor}
\end{align}
Note that $\mathbf{I}_{raw}$ is the original camera RAW image (not the output of RGB2RAW network) because our objective here is to first learn RAW to sRGB mapping, independently. 
\vspace{0.4em}\\
\noindent \textbf{Color attention unit.}
To train the CycleISP, we use the MIT-Adobe FiveK dataset \cite{mit_fivek} that contains images from several different cameras having diverse and complex ISP systems. 
It is extremely difficult for a CNN to accurately learn a RAW to sRGB mapping function for all different types of cameras (as one RAW image can potentially map to many sRGB images). 
One solution is to train one network for each camera ISP \cite{Chen2018,Schwartz2018,zamir2019ISP}. 
However, such solutions are not  
scalable and the performance may not generalize to other cameras. 
To address this issue, we propose to include a color attention unit in the RAW2RGB network that provides explicit color attention via a \emph{color correction branch}.

The color correction branch is a CNN that takes as input an sRGB image $\mathbf{I}_{rgb}$ and generates a color-encoded deep feature tensor $T_{color}\in \mathbb{R}^{\frac{H}{2}{\times}\frac{W}{2}{\times} C}$. 
In the color correction branch, we first apply Gaussian blur to $\mathbf{I}_{rgb}$, followed by a convolutional layer $M_3$, two RRGs and a gating mechanism with sigmoid activation $\sigma$:
\begin{align}
T_{color} = \sigma(M_4(RRG_2(RRG_1( M_3(K\ast \mathbf{I}_{rgb}) )))),
\label{Eq:ccb}
\end{align}
where $\ast$ denotes convolution, and $K$ is the Gaussian kernel with standard deviation empirically set to 12. 
This strong blurring operation ensures that only the color information flows through this branch, whereas the structural content and fine texture comes from the main RAW2RGB network. 
Using weaker blurring will undermine the effectiveness of the feature tensor $T_{d'}$ of Eq.~(\ref{Eq:raw2rgb DF tensor}). 
The overall color attention unit process becomes:
\begin{align}
T_{atten} = T_{d'} + (T_{d'} \otimes T_{color}),
\label{Eq:ccb unit}
\end{align}
where, $\otimes $ is Hadamard product. 
To obtain the final sRGB image $\hat{\mathbf{I}}_{rgb}$, the output features $T_{atten}$ from the color attention unit are passed through a RRG module, a convolutional layer $M_4$ and an upscaling layer $M_{up}$ \cite{Shi2016}, respectively:
\begin{align}
\hat{\mathbf{I}}_{rgb} = M_{up}(M_5(RRG_K(T_{atten}))).
\label{Eq:raw2rgb}
\end{align}
For optimizing RAW2RGB network, we use the $L_1$ loss:
\begin{equation}
\label{Eq:loss raw2rgb}
\mathcal{L}_{{r\rightarrow s}}(\hat{\mathbf{I}}_{rgb},\mathbf{I}_{rgb})
=
\left \| \hat{\mathbf{I}}_{rgb} - \mathbf{I}_{rgb}  \right \|_{1} .
\end{equation}

\begin{figure}[t!]
\begin{center}
 \includegraphics[width=0.85\linewidth]{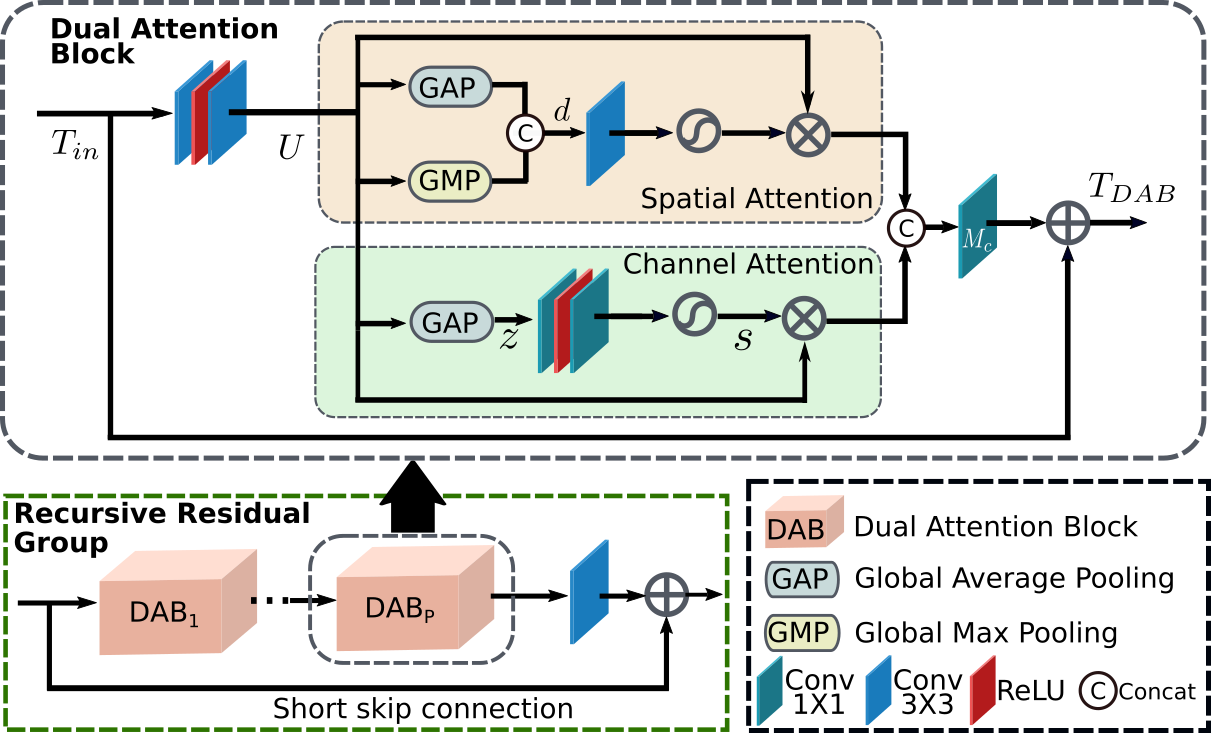}  
\end{center}\vspace{-1.4em}
    \caption{Recursive residual group (RRG) contains multiple dual attention blocks (DAB). Each DAB contains spatial attention and channel attention modules.}\vspace{-0.5em}
    \label{Fig:rrg}
\vspace{-2mm}
\end{figure}

\subsection{RRG: Recursive Residual Group}
\label{sec:rrg}
Motivated by the advances of recent low-level vision methods \cite{ren2019rain,Zhang2018dehaze,DnCNN,zhang2018rcan} based on the residual learning framework \cite{He2016},
we propose the RRG module, as shown in Fig.~\ref{Fig:rrg}. 
The RRG contains $P$ dual attention blocks (DAB). 
The goal of each DAB is to suppress the less useful features and only allow the propagation of more informative ones. 
The DAB performs this feature recalibration by using two  
attention mechanisms: (1) channel attention (CA) \cite{hu2018squeeze}, and (2) spatial attention (SA) \cite{woo2018cbam}.  
The overall process is:
\begin{align}
T_{DAB} = T_{in} + M_c([\text{CA}(U), \text{SA}(U)]),
\label{Eq:dab}
\end{align}
where $U \in \mathbb{R}^{H\times W \times C}$ denotes features maps that are obtained by applying two convolutions on input tensor $T_{in} \in \mathbb{R}^{H\times W \times C}$ at the beginning of the DAB, and $M_c$ is the last convolutional layer with filter size $1\times1$. 
\vspace{0.4em}\\
\noindent \textbf{Channel attention.} This branch is designed to exploit the inter-channel dependencies of convolutional features.
It first performs a \emph{squeeze} operation in order to encode the spatially global context, which is then followed by an \emph{excitation} operation to fully capture channel-wise relationships  \cite{hu2018squeeze}. 
The squeeze operation is realized by applying global average pooling (GAP) on feature maps $U$, thus yielding a descriptor $z \in \mathbb{R}^{1\times 1 \times C}$. 
The excitation operator recalibrates the descriptor $z$ using two convolutional layers followed by the sigmoid activation and results in activations $s \in \mathbb{R}^{1\times 1 \times C}$. 
Finally, the output of CA branch is obtained by rescaling $U$ with the activations $s$.
\vspace{0.4em}\\
\noindent \textbf{Spatial attention.} This branch exploits the inter-spatial relationships of features and computes a spatial attention map that is then used to rescale the incoming features $U$. 
To generate the spatial attention map, we first independently apply global average pooling and max pooling operations on features $U$ along the channel dimensions and concatenate the output maps to form a spatial feature descriptor $d \in \mathbb{R}^{H\times W \times 2}$. 
This is followed by a convolution and sigmoid activation to obtain the spatial attention map. 

\subsection{Joint Fine-tuning of CycleISP}
\label{sec:finetune}
Since the RGB2RAW and RAW2RGB networks are initially trained independently, they may not provide the optimal-quality images due to the disconnection between them. 
Therefore, we perform joint fine-tuning in which the output of RGB2RAW becomes the input of RAW2RGB. 
The loss function for the joint optimization is:
\begin{align}
\label{Eq:loss joint}
\mathcal{L}_{joint}= \beta  \mathcal{L}_{{s\rightarrow r}}(\hat{\mathbf{I}}_{raw},\mathbf{I}_{raw}) + (1{-}\beta)  \mathcal{L}_{{r\rightarrow s}}(\hat{\mathbf{I}}_{rgb},\mathbf{I}_{rgb}), 
\nonumber
\end{align}
where $\beta$ is a positive constant. 
Note that the RAW2RGB network receives gradients from the RAW2RGB sub-loss (only the second term). 
Whereas, the RGB2RAW network receives gradients from both sub-losses, thereby effectively contributing to the reconstruction of the final sRGB image.

\begin{figure}[t!]
\begin{center}
 \includegraphics[width=\linewidth]{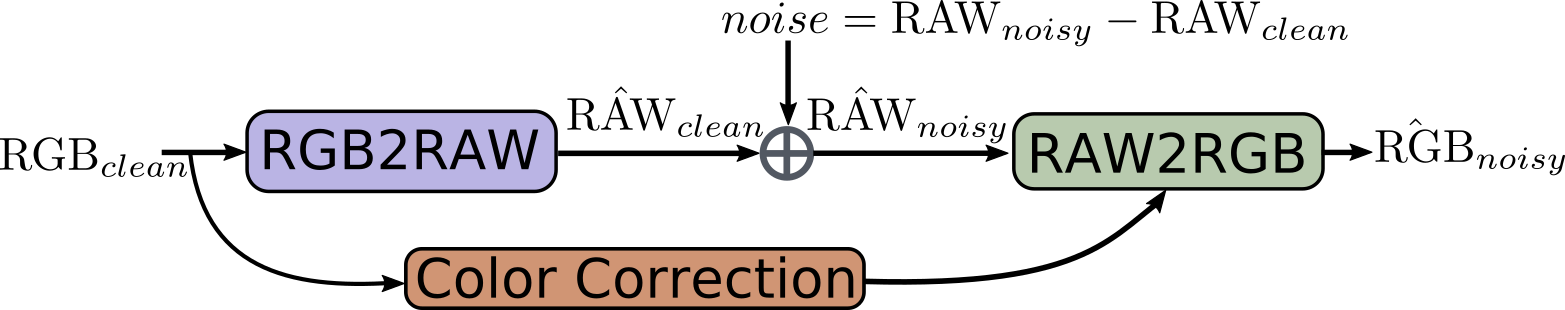}  
\end{center}\vspace{-1.4em}
    \caption{Fine-tuning CycleISP to synthesize realistic sRGB noise data.}\vspace{-2mm}
    \label{Fig:finetune}
\end{figure}

\section{Synthetic Realistic Noise Data Generation}
\label{sec:finetune noisy}
Capturing perfectly-aligned real noise data pairs 
is extremely difficult. 
Consequently, image denoising is mostly studied in artificial settings where Gaussian noise is added to the clean images. 
While the state-of-the-art image denoising methods \cite{MLP,DnCNN} have shown promising performance on these synthetic datasets, they do not perform well when applied on real camera images \cite{sidd,dnd}. 
This is because the synthetic noise data differs fundamentally from real camera data. 
In this section, we describe the process of synthesizing realistic noise image pairs for denoising both in RAW and sRGB space using the proposed CycleISP method. 
\vspace{0.4em}\\
\noindent \textbf{Data for RAW denoising.}
The RGB2RAW network branch of the CycleISP method takes as input a clean sRGB image and converts it to a clean RAW image (top branch, Fig.~\ref{Fig:cycleisp}). 
The noise injection module, which we kept off while training CycleISP, is now turned to the `ON' state. 
The noise injection module adds shot and read noise of different levels to the output of RGB2RAW network. 
We use the same procedure for sampling shot/read noise factors as in \cite{Brooks2019}. 
As such,  we can generate clean and its corresponding noisy image pairs \{$\text{RAW}_{clean}$, $\text{RAW}_{noisy}$\} from any sRGB image.   
\vspace{0.4em}\\
\noindent \textbf{Data for sRGB denoising.}
Given a synthetic $\text{RAW}_{noisy}$ image as input, the RAW2RGB network maps it to a noisy sRGB image (bottom branch, Fig.~\ref{Fig:cycleisp}); hence we are able to generate an image pair \{$\text{sRGB}_{clean}$,$\text{sRGB}_{noisy}$\} for the sRGB denoising problem. 
While these synthetic image pairs are already adequate for training the denoising networks, we can further improve their quality with the following procedure. 
We fine-tune the CycleISP model (Section~\ref{sec:finetune}) using the SIDD dataset \cite{sidd} that is captured with real cameras. 
For each static scene, SIDD contains clean and noisy image pairs in both RAW and sRGB spaces.  
The fine-tuning process is shown in Fig.~\ref{Fig:finetune}. 
Notice that the noise injection module which adds random noise is replaced by (only for fine-tuning) per-pixel noise residue that is obtained by subtracting the real $\text{RAW}_{clean}$ image from the real $\text{RAW}_{noisy}$ image. 
Once the fine-tuning procedure is complete, we can synthesize realistic noisy images by feeding clean sRGB images to the CycleISP model.

\begin{figure}[t!]
\begin{center}
 \includegraphics[width=\linewidth]{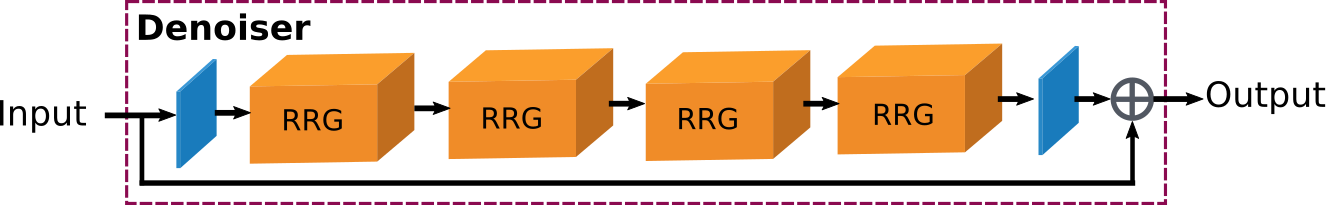}  
\end{center}\vspace{-1.4em}
    \caption{Proposed denoising network. It has the same network structure for denoising both RAW images and sRGB images, except in the handling of input and output.}\vspace{-1.5em}
    \label{Fig:denoising network}
\end{figure}

\section{Denoising Architecture}
As illustrated in Fig.~\ref{Fig:denoising network}, we propose an image denoising network by employing multiple RRGs. 
Our aim is to  apply the proposed network in two different settings: (1) denoising RAW images, and (2) denoising sRGB data. We use the same network structure under both settings, with the only difference being in the handling of input and output. 
For denoising in the sRGB space, the input and output of the network are the 3-channel sRGB images. 
For denoising the RAW images, our network takes as input a 4-channel noisy packed image concatenated with a 4-channel noise level map, and provides us with a 4-channel packed denoised output. 
The noise level map provides an estimate of the standard deviation of noise present in the input image, based on its shot and read noise parameters \cite{Brooks2019}. 

\section{Experiments}

\subsection{Real Image Datasets}
\noindent \textbf{DND \cite{dnd}.} This dataset consists of $50$ pairs of noisy and (nearly) noise-free images captured with four consumer cameras.  
Since the images are of very high-resolution, the providers extract $20$ crops of size $512\times512$ from each image, thus yielding a total of $1000$ patches. 
The complete dataset is used for testing because the ground-truth noise-free images are not publicly available. 
The data is provided for two evaluation tracks: RAW space and sRGB space. 
Quantitative evaluation in terms of PSNR and SSIM can only be performed through an online server \cite{dndwebsite}. 

\vspace{0.4em}\noindent \textbf{SIDD \cite{sidd}.} Due to the small sensor size and high-resolution, smartphone images are much more noisy than those of DSLRs. 
This dataset is collected using five smartphone cameras. 
There are $320$ image pairs available for training and $1280$ image pairs for validation. 
This dataset also provides images both in RAW format and in sRGB space.

\subsection{Implementation Details}
All the models presented in this paper are trained with Adam optimizer ($\beta_1 = 0.9$, and $\beta_2=0.999$) and image crops of $128\times128$. 
Using the Bayer unification and augmentation technique \cite{liu2019learning}, we randomly perform horizontal and vertical flips. 
We set a filter size of $3\times3$ for all convolutional layers of the DAB except the last for which we use $1\times1$. %
\vspace{0.4em}\\
\noindent \textbf{Initial training of CycleISP.} To train the CycleISP model, we use the MIT-Adobe FiveK dataset \cite{mit_fivek}, which contains $5000$ RAW images. 
We process these RAW images using the LibRaw library and generate sRGB images. 
From this dataset,
$4850$ images are used for training and $150$ for validation. 
We use $3$ RRGs and $5$ DABs for both RGB2RAW and RAW2RGB networks, and $2$ RRGs and $3$ DABs for the color correction network. 
The RGB2RAW and RAW2RGB branches of CycleISP are independently trained for $1200$ epochs with a batch size of 4. 
The initial learning rate is $10^{-4}$, which is decreased to $10^{-5}$ after $800$ epochs.

\begin{table}[t]
\begin{center}
\caption{RAW denoising results on the DND benchmark dataset \cite{dnd}. * denotes that these methods use variance stabilizing transform (VST) \cite{makitalo2012VST} to provide their best results. }
\label{Table: raw dnd}
\vspace{-2mm}
\setlength{\tabcolsep}{12pt}
\scalebox{0.70}{
\begin{tabular}{l c c p{1mm} c c }
\toprule
\rowcolor{color3} & \multicolumn{2}{c}{RAW} & & \multicolumn{2}{c}{sRGB} \\
\cline{2-3} \cline{5-6}
  \rowcolor{color3} Method & PSNR~$\textcolor{black}{\uparrow}$ & SSIM~$\textcolor{black}{\uparrow}$ & & PSNR~$\textcolor{black}{\uparrow}$ & SSIM~$\textcolor{black}{\uparrow}$\\
\midrule
TNRD* \cite{TNRD} &   45.70  &  0.96   &  &  36.09  &  0.888 \\
MLP* \cite{MLP}    &   45.71  &  0.963   &  &  36.72  &  0.912 \\
FoE \cite{FoE}     &   45.78  &  0.967   &  &  35.99  &  0.904 \\
EPLL* \cite{EPLL}  &   46.86  &  0.973   &  &  37.46  &  0.925 \\
KSVD* \cite{KSVD}  &   46.87  &  0.972   &  &  37.63  &  0.929 \\
WNNM* \cite{WNNM}  &   47.05  &  0.972   &  &  37.69  &  0.926  \\
NCSR* \cite{NCSR}  &   47.07  &  0.969   &  &  37.79  &  0.923 \\
BM3D* \cite{BM3D}  &   47.15  &  0.974   &  &  37.86  &  0.930 \\
DnCNN \cite{DnCNN}   &   47.37  &  0.976   &  &  38.08  &  0.936 \\
N3Net \cite{N3Net}     &   47.56  &  0.977   &  &  38.32  &  0.938 \\
UPI (Raw) \cite{Brooks2019} &  48.89 &  0.982   &  &  40.17  &  0.962 \\
\midrule
Ours& \textbf{49.13} & \textbf{0.983} & &	\textbf{40.50} & \textbf{0.966}\\
\bottomrule
\end{tabular}}
\end{center}\vspace{-1.4em}
\end{table}

\begin{table}[t]
\begin{center}
\caption{RAW denoising results on the SIDD dataset \cite{sidd}.}
\label{Table: raw sidd}
\vspace{-2mm}
\setlength{\tabcolsep}{12.3pt}
\scalebox{0.70}{
\begin{tabular}{l c c p{0.5mm} c c }
\toprule
\rowcolor{color3} & \multicolumn{2}{c}{RAW} & & \multicolumn{2}{c}{sRGB} \\
\cline{2-3} \cline{5-6}
 \rowcolor{color3} Method & PSNR~$\textcolor{black}{\uparrow}$ & SSIM~$\textcolor{black}{\uparrow}$ & & PSNR~$\textcolor{black}{\uparrow}$ & SSIM~$\textcolor{black}{\uparrow}$\\
\midrule
EPLL \cite{EPLL}   &   40.73  &  0.935   &  &  25.19  &  0.842 \\
GLIDE \cite{GLIDE}  &   41.87  &  0.949   &  &  25.98  &  0.816 \\
TNRD \cite{TNRD}    & 42.77  &  0.945   &  &  26.99  &  0.744 \\
FoE \cite{FoE}          &   43.13  &  0.969   &  &  27.18  &  0.812 \\
MLP \cite{MLP}    &  43.17  &  0.965   &  &  27.52  &  0.788 \\
KSVD \cite{KSVD}   &   43.26  &  0.969   &  & 27.41  &  0.832 \\
DnCNN \cite{DnCNN}  &   43.30  &  0.965   &  &  28.24  &  0.829 \\
NLM \cite{NLM}    &   44.06  &  0.971   &  &  29.39  &  0.846 \\
WNNM \cite{WNNM}   &   44.85  &  0.975   &  &  29.54  & 0.888  \\
BM3D \cite{BM3D}   &   45.52  &  0.980   &  &  30.95  &  0.863 \\
\midrule
Ours & \textbf{52.41} & \textbf{0.993} & &	\textbf{39.47}	&\textbf{ 0.918} \\
\bottomrule
\end{tabular}}
\end{center}\vspace{-1.6em}
\end{table}

\begin{table*}[!t]
\begin{center}
\caption{Denoising sRGB images of the DND benchmark dataset \cite{dnd}.}
\label{Table:srgb dnd}
\vspace{-2mm}
\setlength{\tabcolsep}{8.5pt}
\scalebox{0.70}{
\begin{tabular}{l c c c c c c c c c c c c c c}
\toprule
 \rowcolor{color3} Method & EPLL   & TNRD & NCSR  & MLP   & BM3D & FoE 	& WNNM & KSVD  	& MCWNNM & FFDNet+ & TWSC  & CBDNet & RIDNet  & 
 Ours \\
 
& \cite{EPLL} & \cite{TNRD} & \cite{NCSR} & \cite{MLP} & \cite{BM3D} & \cite{FoE} &  \cite{WNNM}	 & \cite{KSVD} &  \cite{MCWNNM} & \cite{FFDNetPlus} & \cite{TWSC}& \cite{CBDNet} & \cite{RIDNet}  
& \\
\midrule
PSNR~$\textcolor{black}{\uparrow}$ & 33.51 &  33.65 & 34.05 & 34.23 & 34.51 & 34.62 &  34.67 & 36.49 &  37.38 & 37.61 & 37.94 & 38.06 &  39.23 & 
\textbf{39.56}\\
SSIM~$\textcolor{black}{\uparrow}$& 0.824 & 0.831 & 0.835 & 0.833 & 0.851 & 0.885 & 0.865 & 0.898 & 0.929 & 0.942 & 0.940 & 0.942 & 0.953 & 
\textbf{0.956}\\
\bottomrule
\end{tabular}}
\end{center}\vspace{-1.4em}

\end{table*}

\begin{table*}[!t]
\begin{center}
\caption{Denoising sRGB images of the SIDD benchmark dataset \cite{sidd}.}
\label{Table:srgb sidd}
\vspace{-2mm}
\setlength{\tabcolsep}{11pt}
\scalebox{0.70}{
\begin{tabular}{l c c c c c c c c c c c c c c c}
\toprule
 \rowcolor{color3} Method & DnCNN & MLP   & GLIDE & TNRD   & FoE    & BM3D    & WNNM    & NLM    & KSVD    & EPLL    & CBDNet & RIDNet & 
 Ours \\
& \cite{DnCNN} & \cite{MLP} & \cite{GLIDE} & \cite{TNRD} & \cite{FoE} & \cite{BM3D} &  \cite{WNNM}	 & \cite{NLM} &  \cite{KSVD} & \cite{EPLL} & \cite{CBDNet}& \cite{RIDNet} & 
\\
\midrule
PSNR~$\textcolor{black}{\uparrow}$ &  23.66  &   24.71  &    24.71  &  24.73  &  25.58  & 25.65  &   25.78  &   26.76  &  26.88  & 27.11  &  30.78  & 38.71 & 
\textbf{39.52}\\
SSIM~$\textcolor{black}{\uparrow}$ &  0.583 &  0.641 &  0.774 &  0.643 &  0.792 &  0.685 &  0.809 &  0.699 &  0.842 &  0.870 & 0.754   &    0.914 &  
\textbf{0.957}\\
\bottomrule
\end{tabular}}
\end{center}\vspace{-1.4em}
\end{table*}

\begin{figure*}[!t]
\begin{center}
\begin{tabular}[b]{c@{ } c@{ }  c@{ } c@{ } c@{ } c@{ }	}\hspace{-4mm}
    \multirow{4}{*}{\includegraphics[width=.314\textwidth,valign=t]{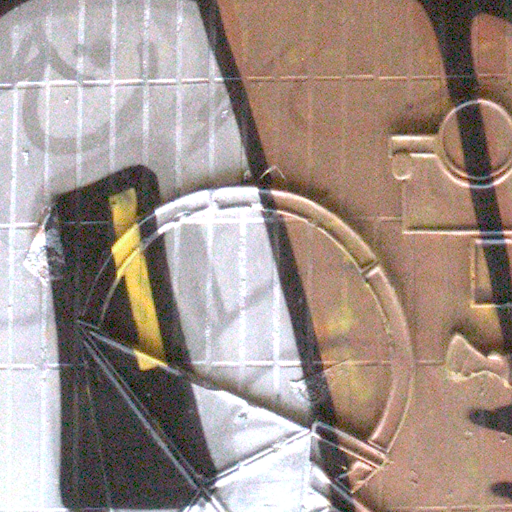}} &   
    \includegraphics[trim={10.8cm 6.5cm  3cm  7.2cm },clip,width=.13\textwidth,valign=t]{Images/DND/RGB/noisy_26_90.png}&
  	\includegraphics[trim={10.8cm 6.5cm  3cm  7.2cm },clip,width=.13\textwidth,valign=t]{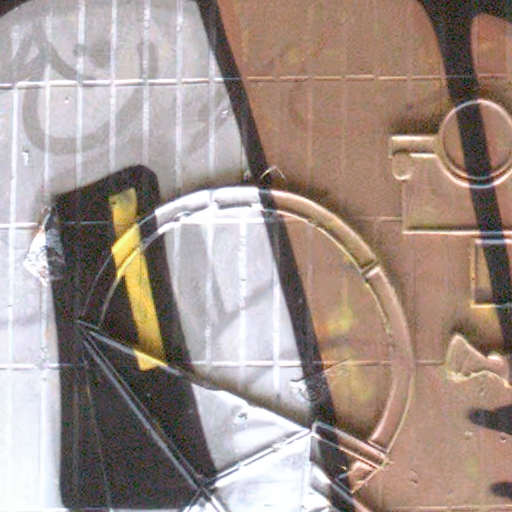}&   
    \includegraphics[trim={10.8cm 6.5cm  3cm  7.2cm },clip,width=.13\textwidth,valign=t]{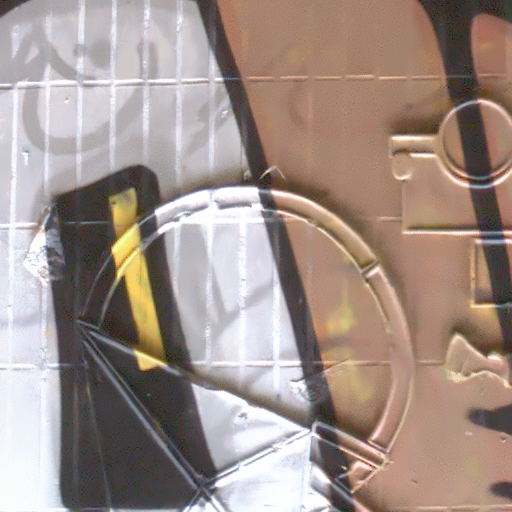}&
      	\includegraphics[trim={10.8cm 6.5cm  3cm  7.2cm },clip,width=.13\textwidth,valign=t]{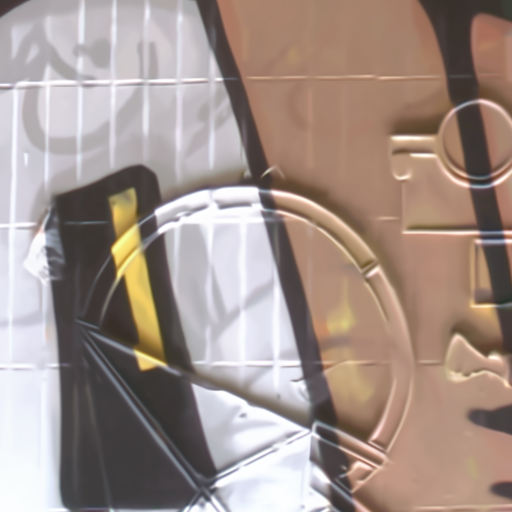}&
      \includegraphics[trim={10.8cm 6.5cm  3cm  7.2cm },clip,width=.13\textwidth,valign=t]{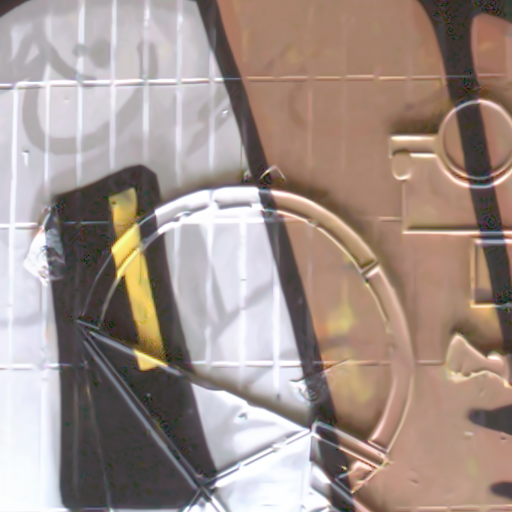}
  
\\
    &  26.90 dB     &30.91 dB  & 32.47 dB & 32.50 dB & 32.74 dB   \\
    & Noisy &BM3D~\cite{BM3D}  & NC~\cite{lebrun2015NC}    & TWSC~\cite{TWSC}      & MCWNNM~\cite{MCWNNM} \\

    &
    \includegraphics[trim={10.8cm 6.5cm  3cm  7.2cm },clip,width=.13\textwidth,valign=t]{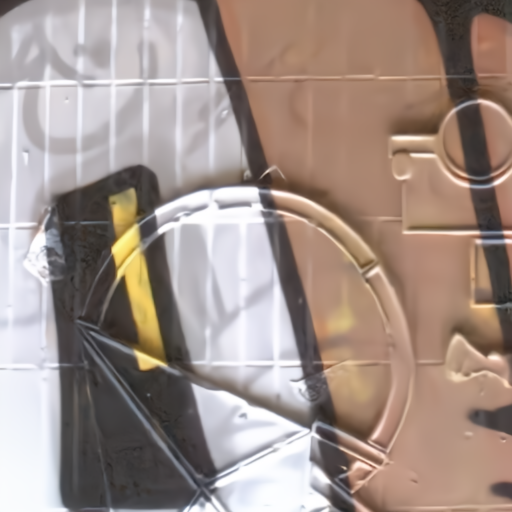}&
    \includegraphics[trim={10.8cm 6.5cm  3cm  7.2cm },clip,width=.13\textwidth,valign=t]{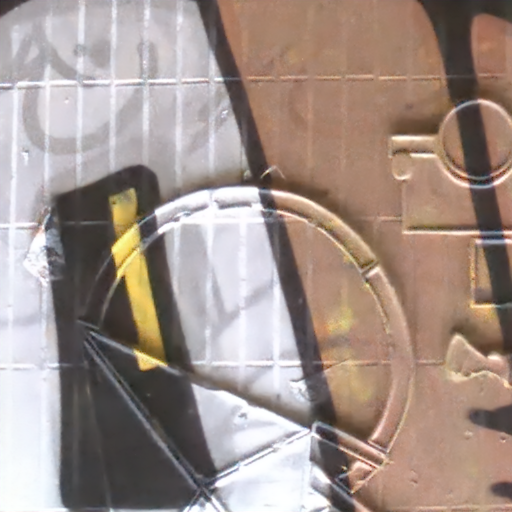}&
    \includegraphics[trim={10.8cm 6.5cm  3cm  7.2cm },clip,width=.13\textwidth,valign=t]{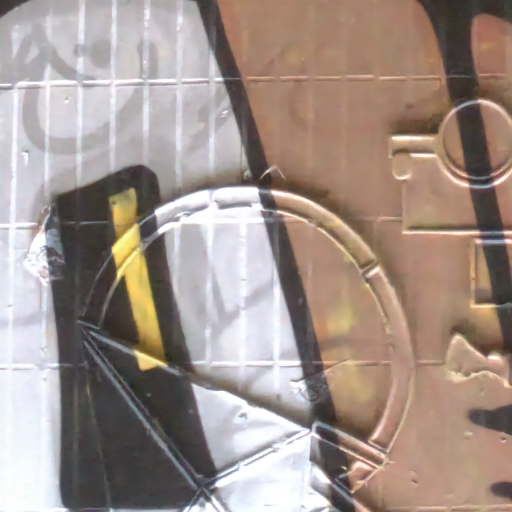}&  
     \includegraphics[trim={10.8cm 6.5cm  3cm  7.2cm },clip,width=.13\textwidth,valign=t]{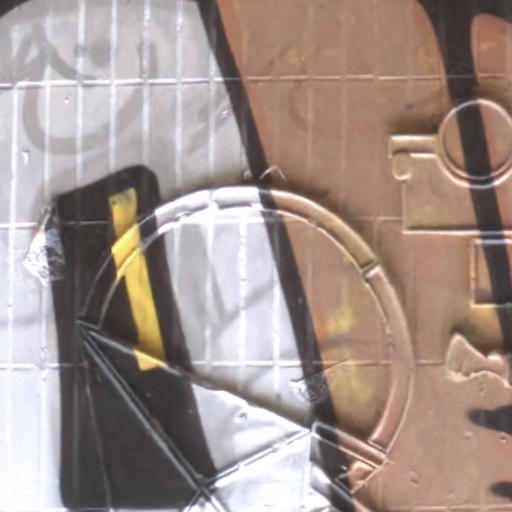}&
     \includegraphics[trim={10.8cm 6.5cm  3cm  7.2cm },clip,width=.13\textwidth,valign=t]{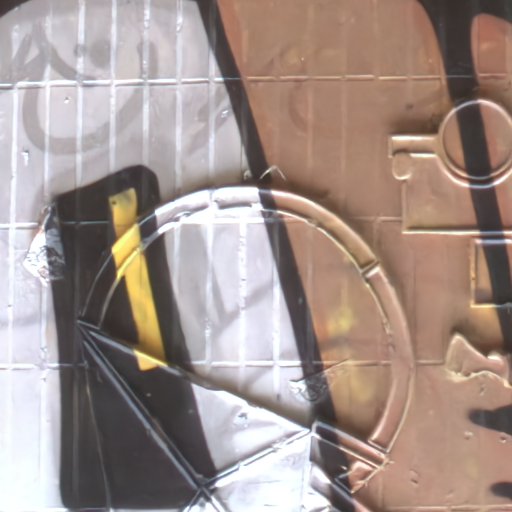}\\

     26.90 dB& 33.05 dB& 33.29 dB & 33.62 dB 
     & 34.09 dB & \textbf{34.32 dB}\\
           Noisy Image  & FDDNet~\cite{FFDNetPlus}&DnCNN~\cite{DnCNN} & CBDNet~\cite{CBDNet}        
           & RIDNet~\cite{RIDNet}     & Ours \\
    
\end{tabular}
\end{center}
\vspace{-6mm}
\caption{Denoising sRGB image from DND~\cite{dnd}. Our method preserves better structural content than other algorithms. }
\label{Fig:srgb dnd}
\vspace*{-4mm}
\end{figure*}

\vspace{0.4em}
\noindent \textbf{Fine-tuning CycleISP.} This process is performed twice: first with the procedure presented in Section~\ref{sec:finetune}, and then with the method of Section~\ref{sec:finetune noisy}. 
In the former case, the output of the CycleISP model is noise-free, and in the latter case, the output is noisy. 
For each fine-tuning stage, we use $600$ epochs, batch size of $1$ and learning rate of $10^{-5}$.  
\vspace{0.4em}\\
\noindent \textbf{Training denoising networks.} We train four networks to perform denoising on: (1) DND RAW data, (2) DND sRGB images, (3) SIDD RAW data, and (4) SIDD sRGB images. 
For all four networks, we use $4$ RRGs and $8$ DABs, $65$ epochs, batch size of 16, and initial learning rate of $10^{-4}$ which is decreased by a factor of 10 after every $25$ epochs.  
We take $1$ million images from the MIR flickr extended dataset \cite{flickr} and split them into a ratio of 90:5:5 for training, validation and testing. 
All the images are preprocessed with the Gaussian kernel ($\sigma =1$) to reduce the effect of noise, and other artifacts. 
Next, we synthesize clean/noisy paired training data (both for RAW and sRGB denoising) using the procedure described in Section~\ref{sec:finetune noisy}.

\subsection{Results for RAW Denoising}
In this section, we evaluate the denoising results of the proposed CycleISP model with existing state-of-the-art methods on the RAW data from DND~\cite{dnd} and SIDD~\cite{sidd} benchmarks. 
Table~\ref{Table: raw dnd} shows the quantitative results (PSNR/SSIM) of all competing methods on the DND dataset
obtained from the website of the evaluation server \cite{dndwebsite}. 
Note that  there are two super columns in the table listing the values of image quality metrics. 
The numbers in the sRGB super column are provided by the server after passing the denoised RAW images through the camera imaging pipeline \cite{karaimer2016software} using image metadata. 
Our model consistently performs  
better against the learning-based as well as conventional denoising algorithms.
Furthermore, the proposed model has $\sim$5$\times$ lesser parameters than previous best method \cite{Brooks2019}.
The trend is similar for the SIDD dataset, as shown in Table~\ref{Table: raw sidd}. Our algorithm achieves 6.89~dB improvement in PSNR over the BM3D algorithm\cite{BM3D}.   

A visual comparison of our result against the state-of-the-art algorithms is presented in Fig.~\ref{Fig:teaser}. 
Our model is very effective in removing real noise, especially the low-frequency chroma noise and defective pixel noise.

\begin{figure}[t]
\begin{center}
\begin{tabular}[t]{c@{ }c@{ }c@{ }c} \hspace{-2mm}
\includegraphics[width=.11\textwidth]{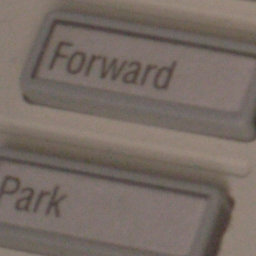}&    \hspace{-1.2mm}
\includegraphics[width=.11\textwidth]{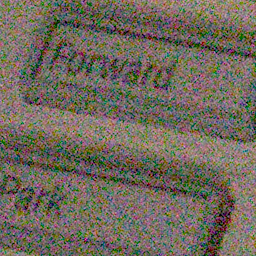}& \hspace{-1.2mm}
\includegraphics[width=.11\textwidth]{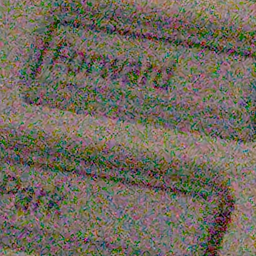}&   \hspace{-1.2mm}
\includegraphics[width=.11\textwidth]{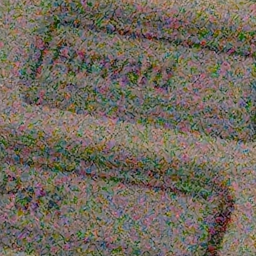}\\
 & 18.25 dB & 19.70 dB& 20.76 dB \\
Reference & Noisy & FFDNet~\cite{FFDNetPlus}& DnCNN~\cite{DnCNN} \\ \hspace{-2mm}
\includegraphics[width=.11\textwidth]{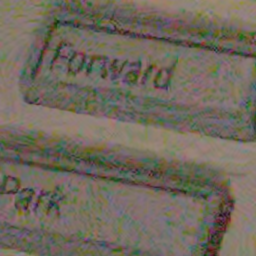}&  \hspace{-1.2mm}
\includegraphics[width=.11\textwidth]{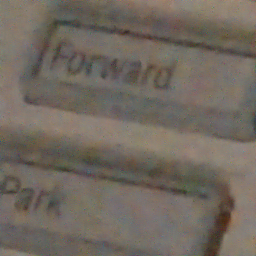}&  \hspace{-1.2mm}
\includegraphics[width=.11\textwidth]{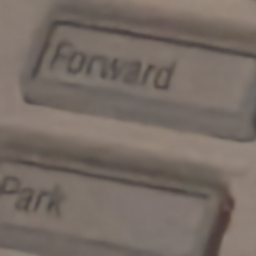}&  \hspace{-1.2mm}
\includegraphics[width=.11\textwidth]{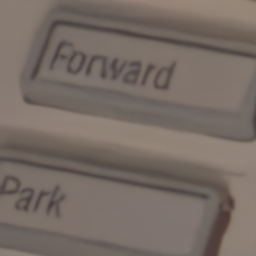}\\
25.75 dB& 28.84 dB & 35.57 dB& \textbf{36.75 dB}\\
BM3D~\cite{BM3D}  & CBDNet~\cite{CBDNet}  & RIDNet~\cite{RIDNet}& Ours \\
\end{tabular}
\end{center}
\vspace*{-6mm}
\caption{Denoising results of different methods on a challenging sRGB image from the SIDD dataset~\cite{sidd}.}
\label{Fig:srgb sidd}
\vspace*{-2mm}
\end{figure}

\subsection{Results for sRGB Denoising}
While it is recommended to apply denoising on RAW data (where noise is uncorrelated and less complex) \cite{ghimpecteanu2016local}, denoising is commonly studied in the sRGB domain. 
We compare the denoising results of different methods on sRGB images from the DND and SIDD datasets. 
Table~\ref{Table:srgb dnd} and~\ref{Table:srgb sidd}
show the scores of image quality metrics. 
Overall, the proposed model performs favorably against 
the state-of-the-art. Compared to the recent best algorithm RIDNet~\cite{RIDNet}, our approach demonstrates the performance gain of 0.33~dB and 0.81~dB on DND and SIDD datasets, respectively.

Fig.~\ref{Fig:srgb dnd} and~\ref{Fig:srgb sidd} illustrate the sRGB denoising results on DND and SIDD, respectively.
To remove noise, most of the evaluated algorithms either produce over-smooth images (and sacrifice image details) or generate images with splotchy texture and chroma artifacts.
In contrast, our method generates clean and artifact-free results, while faithfully preserving image details.

\subsection{Generalization Test}
To compare the generalization capability of the denoising model trained on the synthetic data generated by our method and that of \cite{Brooks2019}, we perform the following experiments. 
We take the (publicly available) denoising model of \cite{Brooks2019} trained for DND, and directly evaluate it on the RAW images from the SIDD dataset. 
We repeat the same procedure for our denoising model as well.
For a fair comparison, we use the same network architecture (U-Net) and noise model as of \cite{Brooks2019}. 
The only difference is data conversion from sRGB to RAW. 
The results in Table~\ref{Table:generalization} show that the denoising network trained with our method not only performs well on the DND dataset but also shows promising generalization to the SIDD set (a gain of $\sim$ 1~dB over \cite{Brooks2019}).

\begin{table}[t]
\begin{center}
\caption{Generalization Test. U-Net model is trained only for DND \cite{dnd} with our technique and with the UPI \cite{Brooks2019} method, and directly evaluated on the SIDD dataset \cite{sidd}.}\vspace{-2.5mm}
\label{Table:generalization}
\setlength{\tabcolsep}{16.5pt}
\scalebox{0.70}{
\begin{tabular}{l c c p{1mm} c c }
\toprule
\rowcolor{color3} & \multicolumn{2}{c}{DND \cite{dnd}} & & \multicolumn{2}{c}{SIDD \cite{sidd}} \\
\cline{2-3} \cline{5-6}
 \rowcolor{color3} Method & PSNR & SSIM & & PSNR & SSIM  \\
\midrule
UPI \cite{Brooks2019}   &   48.89  & 0.9824   &  &  49.17  &  0.9741 \\
Ours       &   \textbf{49.00}  & \textbf{0.9827}  &  & \textbf{50.14}  & \textbf{0.9758} \\
\bottomrule
\end{tabular}}
\end{center}\vspace{-1.4em}
\end{table}

\subsection{Ablations}
We study the impact of individual contributions by progressively integrating them to our model. 
To this end, we use the RAW2RGB network that maps \emph{clean} RAW image to \emph{clean} sRGB image. 
Table~\ref{table:ablation raw2rgb} shows that the skip connections cause the largest performance drop, followed by the color correction branch. 
Furthermore, it is evident that the presence of both CA and SA is important, as well as their configuration (see Table~\ref{table:ablation layout}), for the overall performance. 

\begin{table}[t]
\centering
\caption{Ablation study: RAW2RGB branch.}
\label{table:ablation raw2rgb}
\vspace{-2.5mm}
\setlength{\tabcolsep}{7.3pt}
\scalebox{0.70}{
\begin{tabular}{l c c c c c c}
\toprule
Short skip connections   &      & \ch&\ch   & \ch &  \ch &   \ch  \\ 
Color correction branch  & \ch  & \ch& 	    & \ch  &  \ch &  \ch  \\  
Channel Attention (CA)          & \ch  &  &\ch   & \ch  &       &  \ch  \\
Spatial attention (SA)         &  \ch  &  &\ch    & 	  &   \ch     &  \ch \\\midrule
 PSNR (in dB)   & 23.22 & 42.96 & 33.58 & 44.67 & 45.08 & \textbf{45.41} \\  \bottomrule
\end{tabular}}
\vspace*{-1.6mm}
\end{table}

\begin{table}[t]
\centering
\caption{Layout of SA and CA in DAB.}
\label{table:ablation layout}
\vspace{-2.5mm}
\setlength{\tabcolsep}{15pt}
\scalebox{0.70}{
\begin{tabular}{l c c c}
\toprule \rowcolor{color3}  Layout & CA + SA & SA + CA & CA \& SA in parallel \\ \midrule
PSNR (in dB)&   45.17  & 45.16 & \textbf{45.41} \\ \bottomrule
\end{tabular}}
\vspace*{-1.6mm}
\end{table}

\subsection{Color Matching For Stereoscopic Cinema}

In professional 3D cinema, stereo pairs for each frame are acquired using a stereo camera setup, with two cameras mounted on a rig either side-by-side or (more commonly) in a beam splitter formation \cite{Bertalmio2014}.
During movie production, meticulous efforts are required to ensure that the twin cameras perform in exactly the same manner. 
However, oftentimes visible color discrepancies between the two views are inevitable because of the imperfect camera adjustments and impossibility of manufacturing identical lens systems. 
In movie post-production, color mismatch is corrected by a skilled technician, which is an expensive and highly involved procedure \cite{mendiburu2009}.  

\begin{figure}[t]
\begin{center}
 \includegraphics[width=0.85\linewidth]{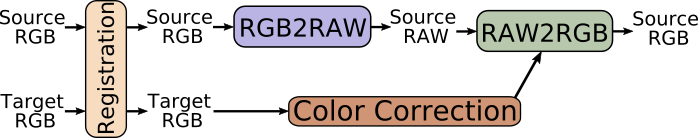}  
\end{center}\vspace{-1.5em}
    \caption{Scheme for color matching 3D pairs. }\vspace{-1.5mm}
    \label{Fig:3d}
\end{figure}

\begin{figure}[t]
\begin{center}
\scalebox{0.95}{
\begin{tabular}[t]{cc} 
\hspace{-7mm}\includegraphics[width=.235\textwidth]{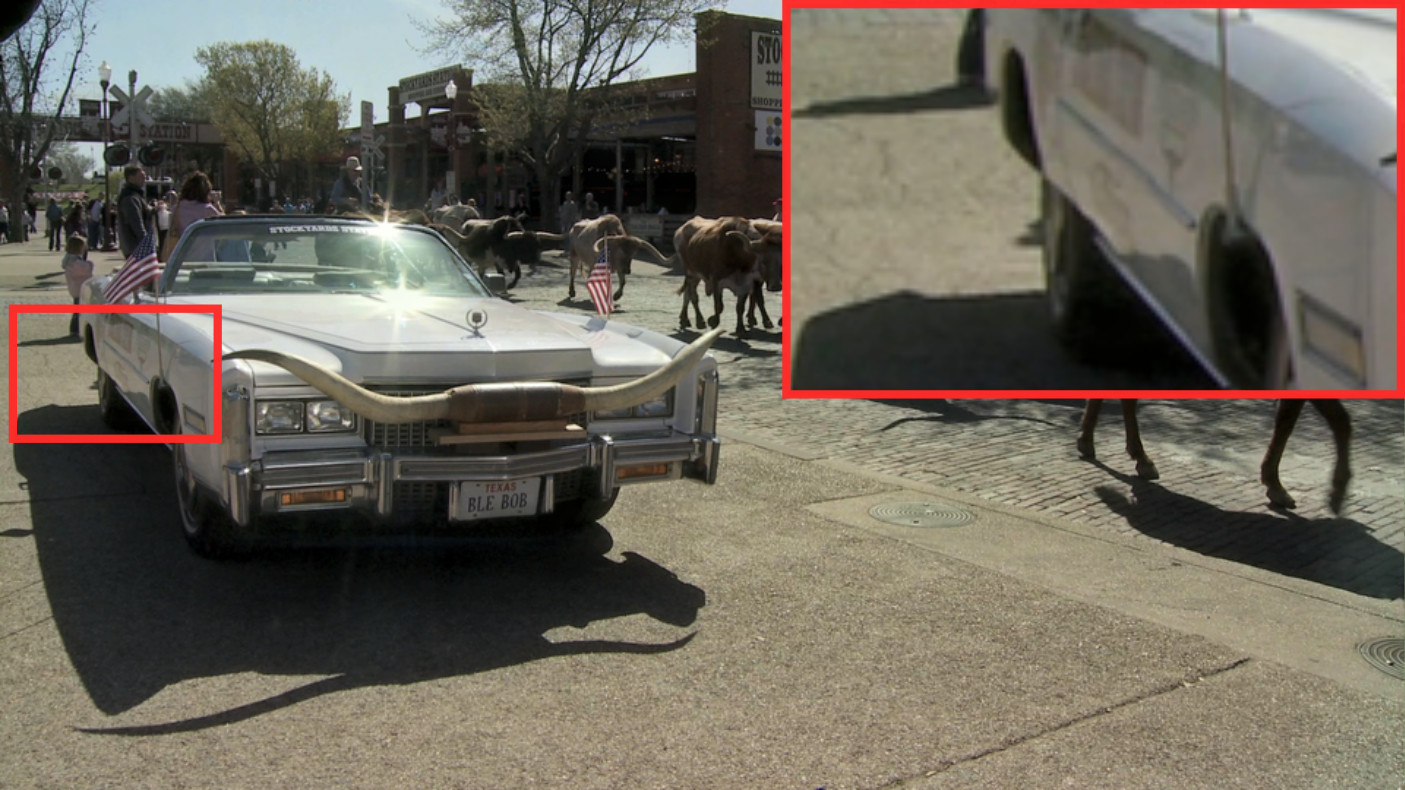}&\hspace{-10mm}
\includegraphics[width=.235\textwidth]{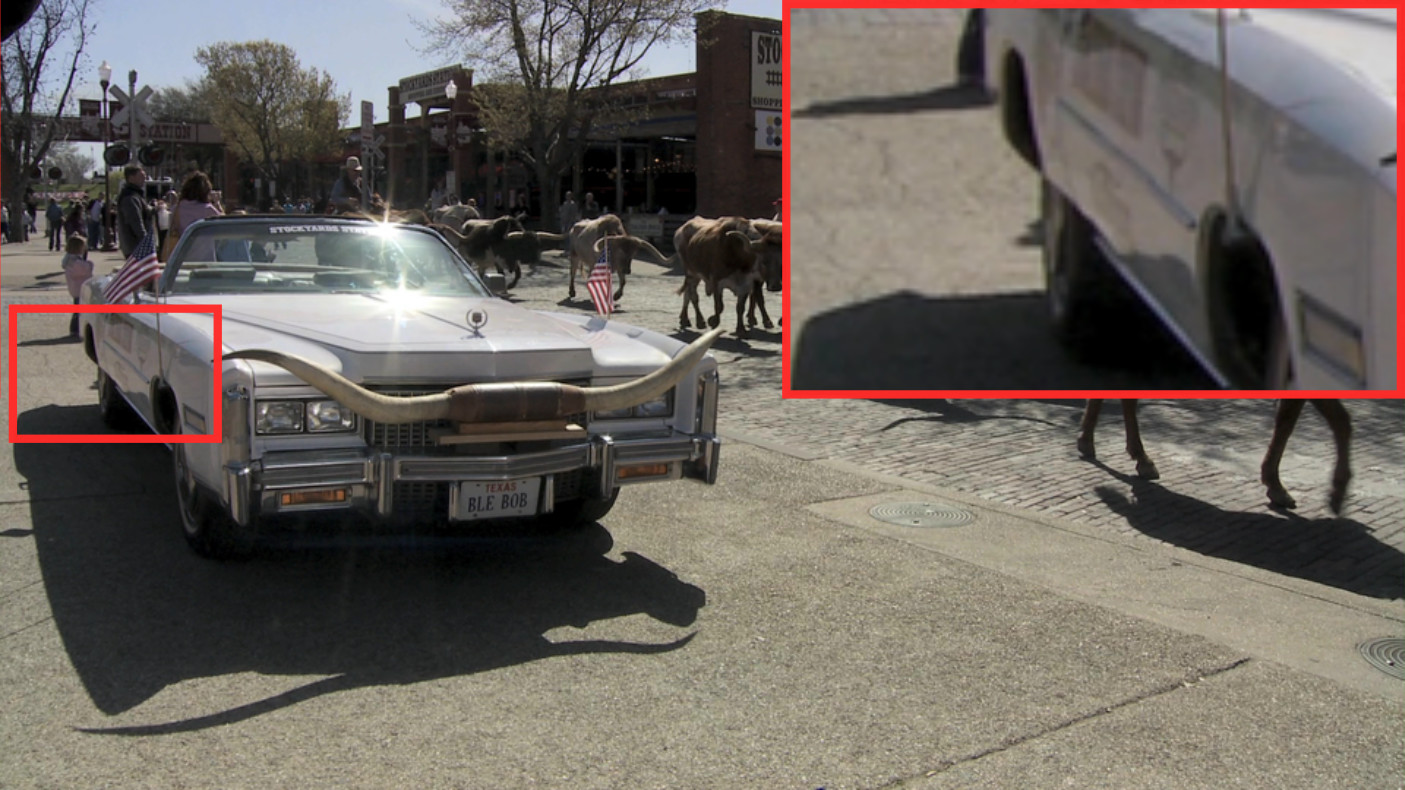}\\
\hspace{-4mm} (a) Target view. (PSNR) & \hspace{-6.5mm}  (b) Source view. 32.17 dB \\
\hspace{-7mm}\includegraphics[width=.235\textwidth]{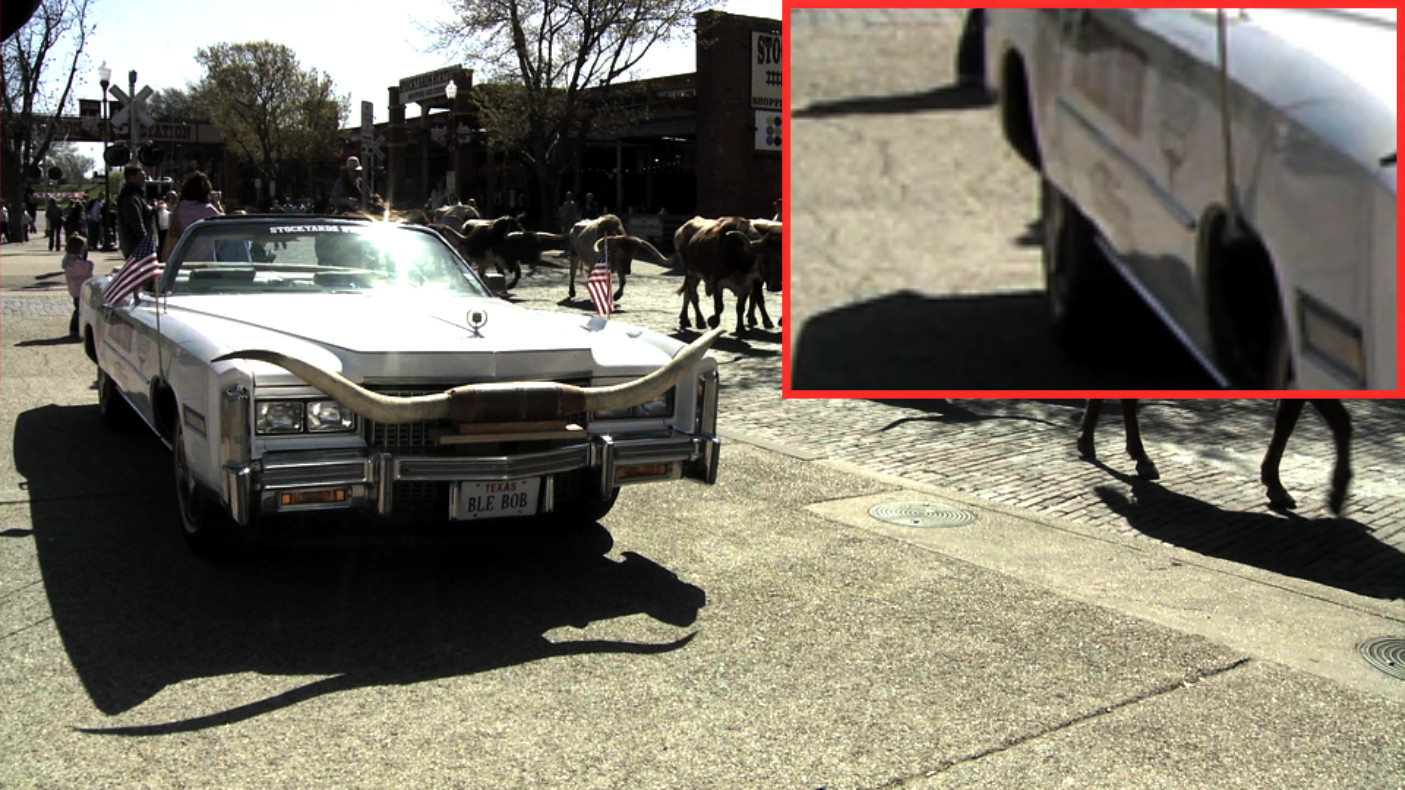}&\hspace{-10mm}
\includegraphics[width=.235\textwidth]{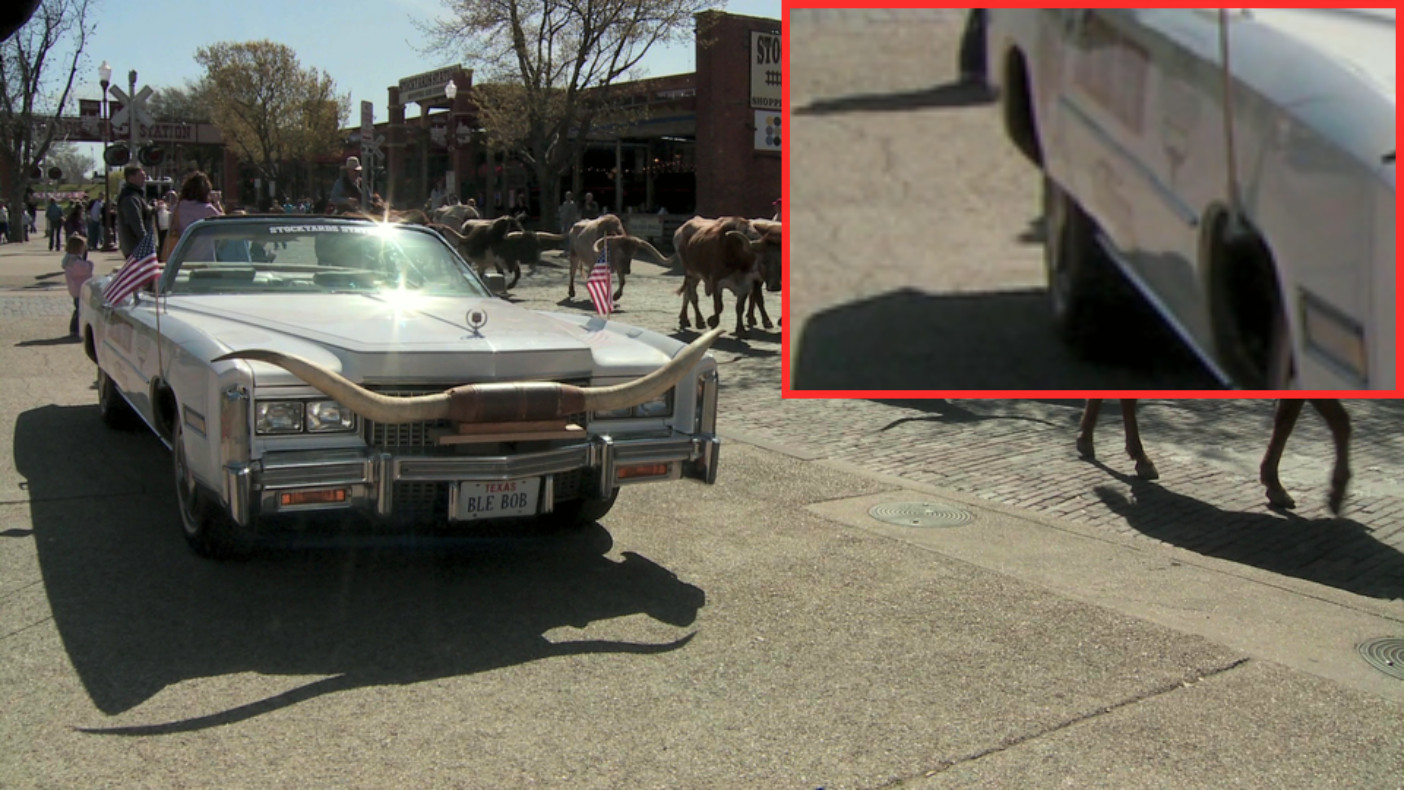}\\
\hspace{-3mm} (c) Reinhard \etal \cite{reinhard2001}. 18.38 dB &  \hspace{-3mm}(d) Kotera \cite{kotera2005}. 32.80 dB \\
\hspace{-7mm}\includegraphics[width=0.235\textwidth]{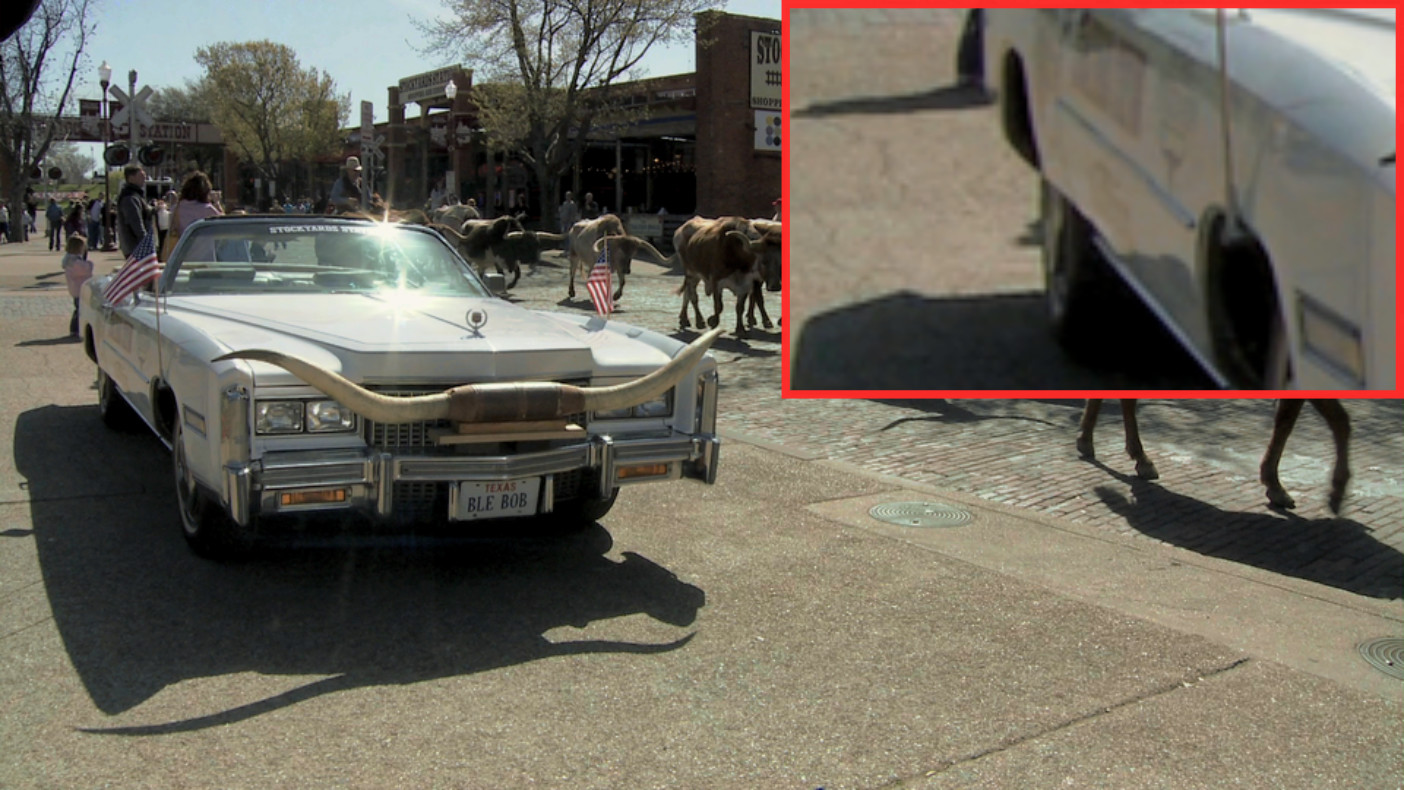}&\hspace{-10mm}
\includegraphics[width=0.235\textwidth]{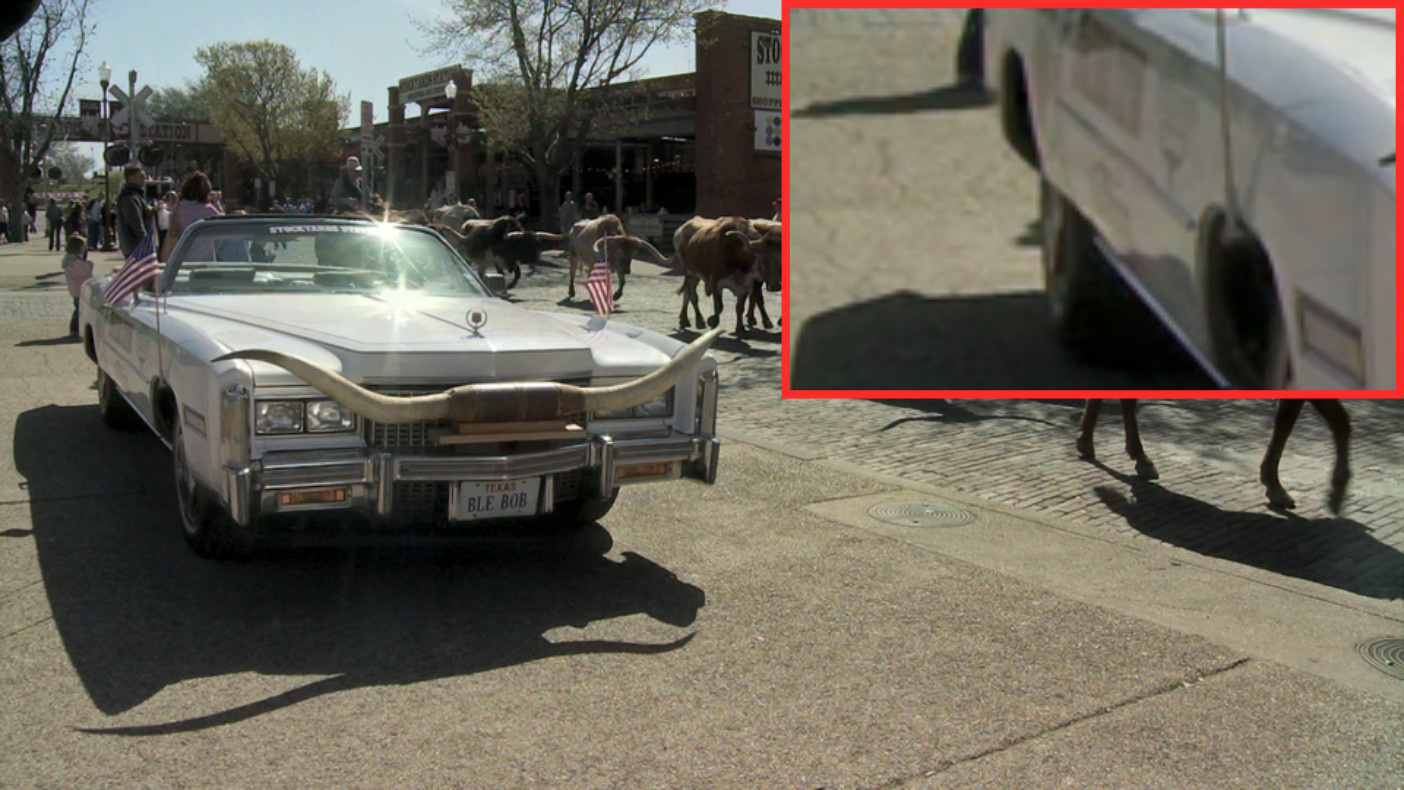}\\
\hspace{-3mm} (e) Piti\'e \etal \cite{pitie2007}. 33.38 dB &  \hspace{-8mm} (f) Ours. \textbf{36.60 dB} \\
\end{tabular}}
\end{center}
\vspace*{-1.4em}
\caption{Example of color correction for 3D cinema. Compare the colors of the ground and side of the car in zoomed-in crops. Images are property of Mammoth HD Inc. }
\label{Fig:3d results}
\vspace*{-3mm}
\end{figure}

With the proposed CycleISP model, we can perform the color matching task, as shown in Fig.~\ref{Fig:3d}. 
Given a stereo pair, we first choose one view as the target and
apply morphing to fully register it with the source view. 
Next, we pass the source RGB image through RGB2RAW model and obtain the source RAW image. 
Finally, we map back the source RAW image to the sRGB space using the RAW2RGB network, but with the color correction branch providing the color information from the `\emph{target}' RGB image (rather than the source RGB).
Fig.~\ref{Fig:3d results} compares our method with three other color matching techniques \cite{kotera2005,pitie2007,reinhard2001}. The proposed method generates results that are perceptually more faithful to the target views than other competing approaches.

\vspace{-0em}

\section{Conclusion}
sIn this work, we propose a data-driven CycleISP framework that is capable of converting sRGB images to RAW data and back to sRGB images.
The CycleISP model allows us to synthesize realistic clean/noisy paired training data both in RAW and sRGB spaces. %
By training a novel network for the tasks of denoising the RAW and sRGB images, we achieve state-of-the-art performance on real noise benchmark datasets (DND~\cite{dnd} and SIDD~\cite{sidd}). 
Furthermore, we demonstrate that the CycleISP model can be applied to the color matching problem in stereoscopic cinema. 
Our future work includes exploring and extending the CycleISP model for other low-level vision problems such as super-resolution and deblurring.

\vspace{0.5em}\noindent\textbf{Acknowledgments.} Ming-Hsuan Yang is supported by the NSF CAREER Grant 149783.

{\small
\bibliographystyle{ieee_fullname}
\bibliography{CycleISP}
}

\end{document}